\documentclass{aa}
\usepackage[]{times}  
\input psfig.sty

\newcommand{\apj}[2]{ApJ #1, #2}

\newcommand{\aeta}[2]{A\&A #1, #2}

\newcommand{\nh}{N$_{\rm H}$}
        
\newcommand{\Halp}{H${\alpha}$}
\newcommand{\He}{\ion{He}{ii} ${\lambda}$4686}
\newcommand{\cn}{\ion{N}{iii}-\ion{C}{iii} ${\lambda \lambda}$4640-60}
\newcommand{\Hbet}{H${\beta}$}

\newcommand{\ergs}{erg s$^{-1}$}

\newcommand{\Msol}{M$_{\odot}$ }

\newcommand{\degree}{\degr}

\newcommand{\rx}{\object{RX\,J0925.7-4758}}
\newcommand{\rxa}{\object{RX\,J0019.8+2156}}
\newcommand{\rxb}{\object{RX\,J0513.9-6951}}
\begin{document}

   \thesaurus{06         
              (
	       13.25.5;  
	       08.23.1;  
	       08.02.1;  
	       08.13.2;  
	       08.09.2;  
	       )}
   \title{The transient jet of the galactic supersoft X-ray source \rx
   \thanks{Based on observations obtained at the European Southern Observatory, 
La Silla (Chile) with the ESO-Danish telescope}
   }

   \author{C. Motch}

   \institute{
              Observatoire Astronomique, UMR 7550 CNRS, 11 rue de l'Universit\'e,
              F-67000 Strasbourg, France     
             }

   \date{Submitted for publication in Astronomy \& Astrophysics, Letters} 

   \offprints{C. Motch}

   \maketitle

   \begin{abstract}

We report the discovery of a transient jet in the supersoft X-ray source \rx . The
jet was observed in the \Halp \ line during a single night in June 1997 and had
disappeared one day later. \rx \ is the third supersoft source to exhibit
collimated outflows. The peak jet velocity of  5,200\,km\,s$^{-1}$, strongly
argues in favour of a white dwarf in \rx. Simple modelling of the jet profile
suggests half opening angles in the range of 17\degree\ to 41\degree \ although the
outflow may be narrower if part of the observed spread in velocity is intrinsic to
the jet. X-ray spectral modelling (Ebisawa et al. \cite{ebisawa96}; Hartmann \&
Heise \cite{hh97}) indicates distances of the order of 10\,kpc or more with the
consequence that \rx \ may be the most optically luminous supersoft source
known.  The overall observational picture points to a massive white dwarf which
may be close to the Chandrasekhar limit.

   \end{abstract}

%

\section{Introduction}

Supersoft X-ray sources are believed to be accreting white dwarfs burning material
on their surface (van den Heuvel et al. \cite{vdh}) in a more or less steady
fashion. Quasi-stable H burning requires high mass transfer rates ($\dot{\rm M}$
$\sim$ 10$^{-7}$ \Msol yr$^{-1}$, Iben  \cite{iben82}). Such high transfer rates
which are at least a factor 10 higher than those occurring in cataclysmic
variables may be achieved by thermally unstable Roche lobe overflow when the mass
donor star is more massive than the accreting object. Steady thermonuclear burning
allows the white dwarf to increase its mass since processed material can remain at
the surface of the accreting object. A large fraction of SNIa could originate in
such systems when the white dwarf mass approaches the Chandrasekhar limit. A
recent review of these systems can be found in Kahabka \& van den Heuvel
(\cite{kavdh97}).

\rx \ and \rxa \ (Beuermann et al. \cite{beuer95}) are the only two luminous and
steady supersoft X-ray sources known so far in the Galaxy. \rx \ was discovered in
the ROSAT all-sky survey and optically identified with a V$\sim$17 heavily reddened
object (Motch et al. \cite{mhp94}). Because of the large interstellar absorption,
only the very high energy part (E $\geq$ 0.5\,keV) of the intrinsically soft energy
distribution is offered to observation. The resulting X-ray spectrum peaks at
0.8\,keV and is unique among supersoft sources (Hartmann \& Heise \cite{hh97}).
Optical photometric and spectroscopic observations suggest an orbital period of
$\sim$ 3.8\,d (Motch \cite{m96}).   

The existence of bipolar jets with velocities in the range of 1,000 - 4,000 km\,
s$^{-1}$ has been recently reported in two supersoft sources, (\rxb : Crampton et
al.  \cite{cramp96}; Southwell et al. \cite{south96} and \rxa : Tomov et al. 
\cite{tomov98}; Becker et al.  \cite{becker98}). In this paper we report on the
discovery of a well formed bipolar outflow with a projected velocity of $\sim$
5,200 km\, s$^{-1}$ in \rx. 

\section{Observations}

\rx \ was observed during two consecutive nights on 1997 June 7 and 8 UT with the
ESO-Danish 1.54\,m telescope and DFOSC equipped with the LORAL-Lesser 2K3EB-C1W7
chip. All spectra have 15\,mn exposure times and on all occasions we used grism \#7
with a slit width of 2.5\arcsec\ in order to accommodate the rather bad seeing
($\sim$ 2\arcsec ) prevailing during the entire run.  In addition to these
spectroscopic data, several 2\,mn long V band exposures were accumulated over the
two nights. During the first night, 3 spectra were obtained in about 1 hour from JD
start times 2,450,606.5499 to 2,450,606.5821. The 5 spectra collected during the
second night span 2.8 hours from JD 2,450,607.4666 to 2,450,607.5823.  

All data were reduced using standard MIDAS procedures. Two dimensional spectra were
calibrated in wavelength using He\,+\,Ne arc lamps exposures closest in time to the
science frame. A S/N optimized procedure extracted one dimensional spectra.  The
spectral range extends from 3,800 to 6,900\,\AA \ with a pixel size of 1.47\,\AA .
Spectral resolution is degraded by the large slit width and bad seeing. FWHM
resolution estimated from the minimum observed width of \Halp \ is about 7.3\,\AA.

\begin{figure}[tbp]
\psfig{figure=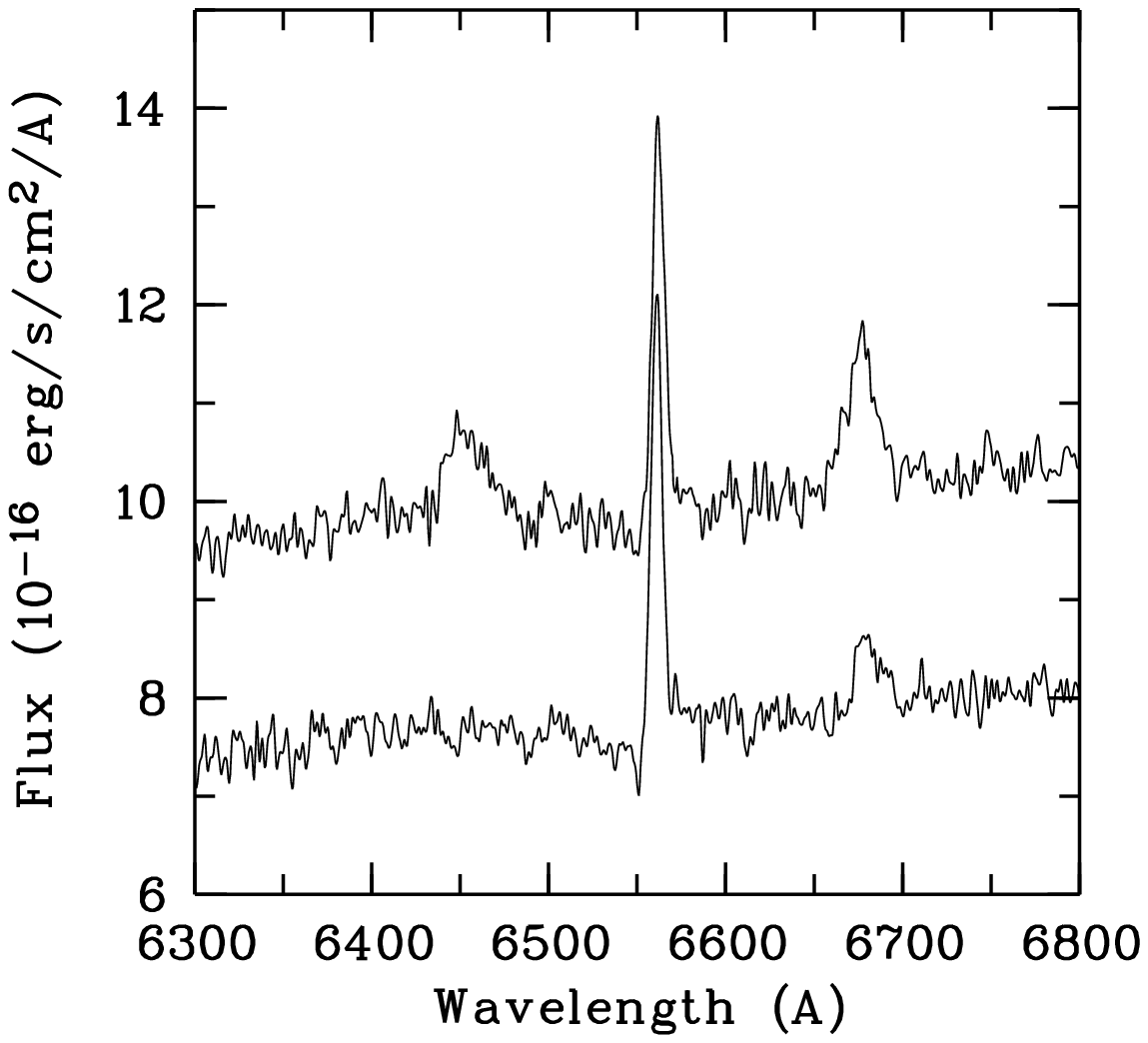,width=8cm,bbllx=1cm,bburx=14cm,bblly=1.5cm,bbury=12.0cm,clip=true}
\psfig{figure=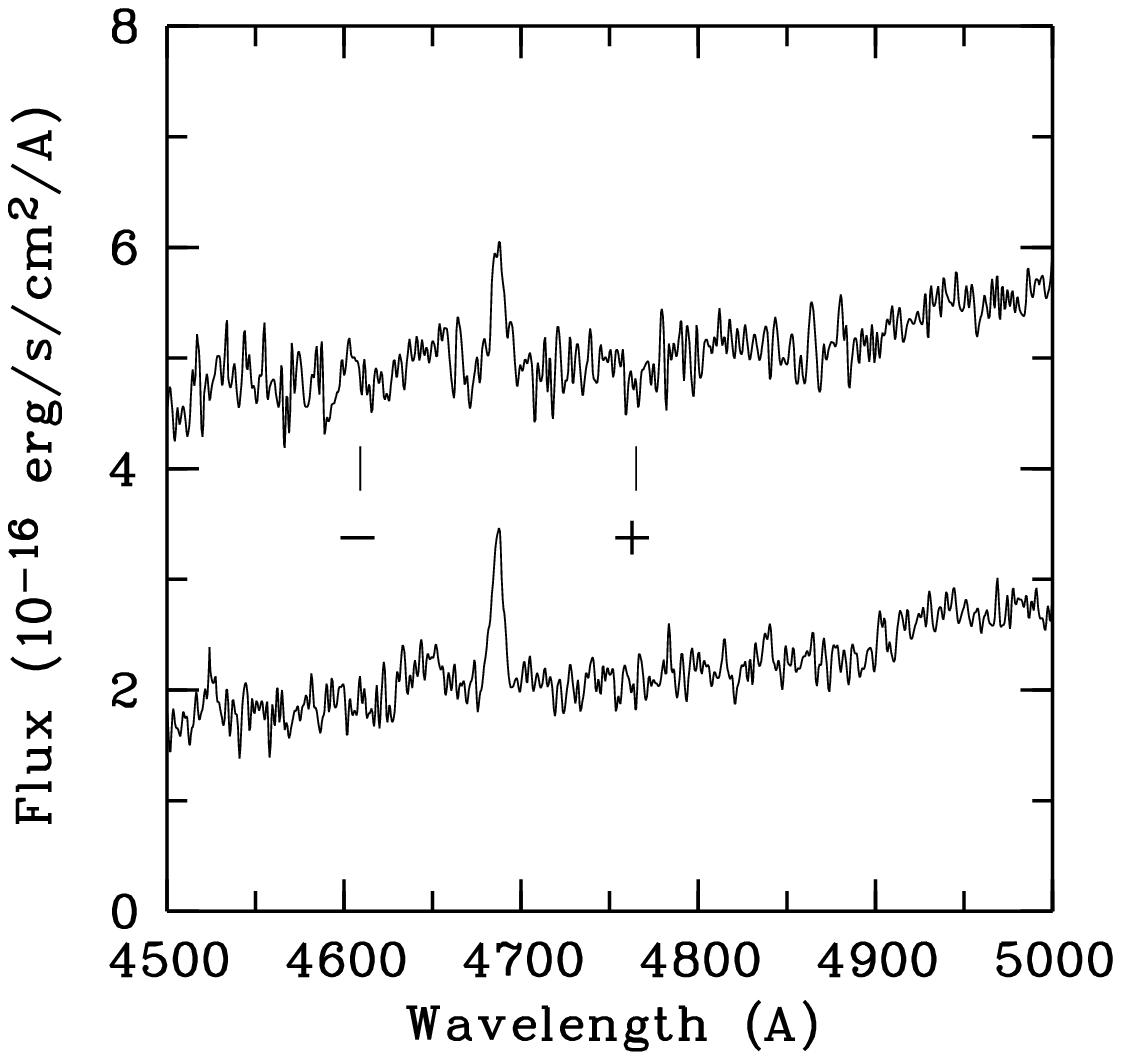,width=8cm,bbllx=1cm,bburx=14cm,bblly=1.5cm,bbury=12.5cm,clip=true}

\caption[]{Mean spectra obtained on June 7 showing the jet feature plotted together
with that obtained on June 8 representing the normal state of the source.  The June 7
spectrum is shifted up by 3 flux units for clarity.  Upper panel: \Halp \ region.
Lower panel: No jet feature is detected around the \He\ emission line.  The \cn\
complex emission is also visible whereas \Hbet \ is not detected}

\label{meanspectra} 
\end{figure}

In Fig. \ref{meanspectra} we plot the average of the three June 7 spectra
exhibiting  the jet feature together with the average of the five June 8 spectra
which are  typical of the normal state of the source. The redshifted \Halp\
component unfortunately coincides in wavelength with the \ion{He}{i}
$\lambda$6678.15 and \ion{He}{ii} $\lambda$6683.2 emission lines. No jet feature is
seen at \ion{He}{ii} $\lambda$4686 which is the second brightest emission line in
the observed wavelength range. Using \Halp \ jet lines as template, we can put an
upper limit of 2.4\,\AA\ to the equivalent width (EW) of the blue \ion{He}{ii}
component. Average spectra of both nights show evidence for P Cyg profile in the
central \Halp \ emission.

\begin{figure}[tbp]
\psfig{figure=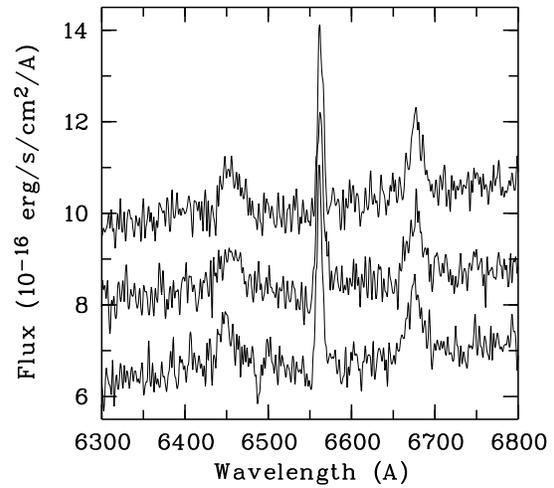,width=8cm,bbllx=1cm,bburx=14cm,bblly=1.5cm,bbury=12.0cm,clip=true}
\caption[]{Individual 15\,mn long spectra obtained on June 7. Spectra are
shifted in flux for clarity. Time goes from bottom to top}

\label{jetspectra}
\end{figure}

Fig. \ref{jetspectra} shows that the jet velocity does not change on a time
scale of one hour. Within the statistical uncertainties, the jet profile
remains constant, showing a clear asymmetric extension towards low absolute
velocities. During the first night, the central \Halp \ line exhibits
shoulders moving in velocity from one spectrum to the other ($\sim$ 20\,mn). Such
variations are not seen during the following night.

\begin{figure}[tbp]
\psfig{figure=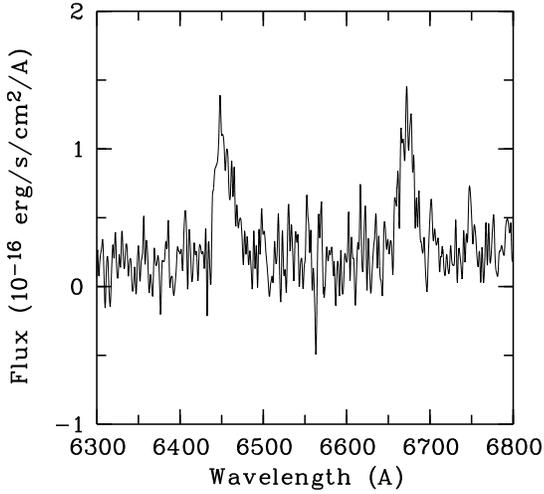,width=8cm,bbllx=1cm,bburx=14cm,bblly=1.5cm,bbury=12.5cm,clip=true}

\caption[]{Jet spectrum obtained by subtracting the average June 8 spectrum
shifted in velocity from the average June 7 one.}

\label{jetonly}
\end{figure}

We tried to correct the red jet component profile for contaminating \ion{He}{i} 
and \ion{He}{ii} lines by subtracting the mean June 8 spectrum shifted by the 44
km\,s$^{-1}$ velocity difference. In the wavelength intervals void of sharp line
features (i.e. excluding the \Halp \ and \ion{He}{i}/\ion{He}{ii} lines), we
smoothed the June 8 continuum with a Gaussian of 7.3\,\AA \ FWHM, equal to the
spectral resolution. This procedure preserves the best statistics and allows
correction for weak water vapour absorption bands which are abundant bluewards of
\Halp . The resulting 'pure' jet spectrum is shown on Fig. \ref{jetonly}. Peak
intensities of both components compare remarkably well. After correction, the
slight asymmetry of the raw red component profile (see Fig. \ref{jetspectra}) seems
to have vanished. This could be due to changing \ion{He}{i}/\ion{He}{ii} line
emission strength between the first and the second night.  Independently of any
correction error due to \ion{He}{i}/\ion{He}{ii} lines, the red and blue component
profiles appear to have different widths, the red profile being about 8\,\AA \
(360\,km\,s$^{-1}$) less extended towards low absolute velocities than the blue
one. The EW of the blue component is 4.1\AA , slightly larger than those of \rxa\
(3\,\AA , Becker et al. \cite{becker98}) and RX\,J0513.9-6951 (2.6\,\AA , Southwell
et al. \cite{south96}). In contrast, the EW of the central \Halp \  (4.9\,\AA) and
\He \ (7.8\,\AA ) lines in \rx \ are significantly smaller than those of other
supersoft sources exhibiting jets. The upper limit on the \He\ to \Halp\ EW ratio
for blueshifted components ($\sim$ 0.6) is compatible with those observed in other
sources and does not suggest a lower excitation level in the jet of \mbox{\rx .}   

The V magnitude of \rx \ was 17.213$\pm$0.016 and 17.133$\pm$0.026 on June 7
and 8 UT respectively. These magnitudes are within the range of those reported
for the source since 1992 (Motch et al. \cite{mhp94}; Motch \cite{m96}) and
there is therefore no evidence that the appearance of the jet is accompanied
by any large change in optical continuum emission.

\section{Jet modelling}

The detailed shape of the shifted \Halp \ lines, in particular their extended wings
towards small absolute velocities potentially contains valuable information on jet
geometry and kinematics. We therefore designed a simple kinematic jet model derived
from that used by Becker et al. (\cite{becker98}). The jet is described as a cone of
half opening angle $\alpha$ in which all atoms move with velocity $v_{j}$. Within
the cone, material is assumed to be flowing uniformly per unit solid angle. In order
to compute line profiles we further assumed that the outflow emission region is
optically thin and convolved the model profile with a Gaussian of FWHM =
333\,km\,s$^{-1}$ ($\sigma$ = 142\,km\,s$^{-1}$) representing the instrumental
profile. 

Since we only have a snapshot observation at an unknown orbital phase, we cannot
constrain the orientation of the jet with respect to orbital plane or with respect
to the axis joining the two stars. In our case, the only relevant angles are
$\alpha$ and the angle $i$ between the line of sight and the jet axis. In this
simplified geometry, the jet axis is aligned with the $z$ axis and the line of
sight is contained in the $x$-$z$ plane. A flow making an angle $\beta$ with
respect to jet axis and an angle $\phi$ with respect to the $x$ axis will have a
projected component \begin{math} V = v_{j}\ (\sin \beta \cos \phi \sin i + \cos
\beta \cos i )\end{math} on the line of sight.

The jet components extend in velocity from about 3,800 to 5,800\,km\,s$^{-1}$ with
a peak at 5,200\,km\,s$^{-1}$. Any effect related to orbital motion is likely to
be  negligible since the K amplitude of the \ion{He}{ii} line is only
$\sim$80\,km\,s$^{-1}$ (Motch \cite{m96}) and since the duration of the observation
($\sim$ 1\,h) is small with respect to the suspected orbital period (P$_{\rm Orb}$
$\sim$ 3.8\,d).

\begin{figure}[tbp]
\psfig{figure=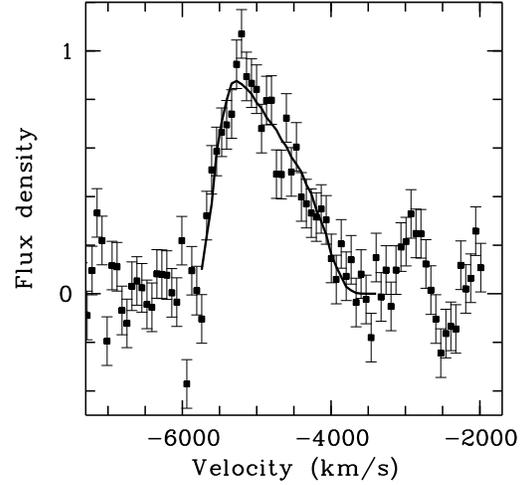,width=8cm,bbllx=1cm,bburx=14cm,bblly=1.5cm,bbury=12.5cm,clip=true}

\caption[]{Blue shifted \Halp \ velocity profile. A best fit model profile ($\chi
^{2}_{32}$ = 38.1) with $\alpha$ = 20.7\degree , $i$ = 24.6\degree, $v_{j}$ =
5,600\,km\,s$^{-1}$ is also shown for comparison}

\label{figfit}
\end{figure}

Considering the uncertainties resulting from the \ion{He}{i}/\ion{He}{ii} line
contamination, we did not try to fit a model jet profile to the redshifted \Halp\
component. The width of the blue component profile and its asymmetry, namely its
larger extension towards low absolute velocities, can be accurately represented by a
well opened jet seen at rather low inclination. As shown on Fig. \ref{figfit}, the fit
is surprisingly good considering the simplicity of the model ($\chi ^{2}_{32}$ =
38.1). At the 99\% confidence level, the formal accepted ranges of inclinations $i$
are 7\degree\ to 13\degree \ and 20\degree\ to 29\degree \  and those of  half-opening
jet angles $\alpha$ are 17\degree\ to 27\degree\ and 34\degree\ to 41\degree . Large
inclination angles correspond to small jet opening angles. The range of $\alpha$
values compares well with that derived for \rxa \ by Becker et al. (\cite{becker98}).
In the framework of this simple model the difference in profile shape between the blue
and red components could reflect different opening angles (assuming inclinations are
identical).

Alternatively, the width and asymmetry of the blueshifted profile could be due to
an intrinsic spread of material velocity in the outflow. This would allow much
narrower opening angles, more consistent with the conception of a jet. In this
picture, the difference in profile shape and extent between the blue and red
components may be for instance interpreted in terms of occultation of the low
velocity part of an accelerating jet by the accretion disc. Finally, line profiles
may also be intrinsically broadened by Keplerian velocity in the inner parts of the
accretion disc (Becker et al. \cite{becker98}).

\section{Discussion and Conclusions}

\rx \ is the third supersoft source in which collimated outflows are observed
demonstrating that jets are common phenomena in this class of high mass transfer
rate accreting binaries.

However, compared to other supersoft sources, the jet of \rx \ appears to be a
rather rare and rapidly variable phenomenon. The jet was detected during only one
among 23 nights of observation performed since the identification of the source in
1992, and it disappeared in less than 24\,h. For comparison, the jet of
RX\,J0513.9-6951 is almost constantly seen at about the same velocity. Crampton et
al. (\cite{cramp96}) report non-detection during only one or two nights. On the
other hand, the jet of \rxa\ is transient on time scales of months (Becker et al.
\cite{becker98}). 

The observation of a jet with a projected velocity of $\sim$ 5,200\,km\,s$^{-1}$
confirms that \rx \ belongs to the class of supersoft sources in spite of its unusual
X-ray spectrum. In particular, if the jet velocity is of the order of the escape velocity
from the central object (see e.g., Livio \cite{livio97}) then the M/R ratio of the
source is similar to that of a white dwarf. Wind velocities of the order of
6,000\,km\,s$^{-1}$ are indeed observed in some cataclysmic variables (Drew  
\cite{drew97})

Fits of NLTE model atmospheres to ASCA data (Ebisawa et al. \cite{ebisawa96}) and 
ROSAT PSPC data (Hartmann \& Heise \cite{hh97}) indicate high effective temperatures
close to 70\,eV and amazingly small source radii in the range of
160-370\,($d$/1\,kpc)\,km. The reduced emitting area has been sometimes considered as
evidence that the source was in fact a neutron star with an extended atmosphere
(Hartmann \& Heise \cite{hh97}, Kylafis \& Xilouris \cite{kx93}). If the jet
originates from the close surrounding of the X-ray emitting surface then the source
radius is of the order of 10$^{9}$\,cm for solar masses objects, independently of the
actual nature of the central engine, white-dwarf or shrouded neutron star. From this
point of view, \rx \ does not look different from other supersoft sources and its
distance should be at least 10\,kpc in order for the X-ray emitting region to reach a
radius similar to that of the jet producing region. At 10\,kpc, the bolometric
luminosity is 5 10$^{37}$ \ergs \  and the radius of the source is at most 3,700\,km
suggesting a very massive white dwarf. However, as the bulk of the energy distribution
of \rx \ is masked by photoelectric absorption, it is still possible that
spectral fits give somewhat biased source parameters.

Optical data do not rule out such large distances. The interstellar absorption towards
the source (\nh\ $\sim$ 1.3 10$^{22}$ cm$^{-2}$, Motch et al. \cite{mhp94}) is similar
to the integrated galactic value while a large part of the reddening probably takes
place rather locally in the Vela sheet molecular cloud located at 425\,pc. At 10\,kpc,
the intrinsic V magnitude of \rx \ is M$_{\rm V}$ = $-$4, two magnitudes brighter than
the brightest of the Magellanic supersoft sources, RX\,J0513.9-6951. This high optical
luminosity could reflect the long orbital period and large accretion disc of \rx . In
supersoft sources, the white dwarf luminosity due to nuclear burning is much larger
than the total accretion luminosity of the disc and X-ray reprocessing in the disc and
on the secondary atmosphere should play an important role (see e.g. Popham \&
DiStefano \cite{pd96}). If as for low-mass X-ray binaries M$_{\rm V}$ scales as
1.67$\times$log(P$_{\rm Orb}$) (van Paradijs and McClintock \cite{vm94}), then the
larger orbital period and accretion disc in \rx \ may already explain a 1.2 magnitude
difference. Different disc rim structures (Meyer-Hofmeister et al. \cite{msm97}) and
uncertainties on A$_{\rm V}$ could account for the rest of the difference in absolute
magnitude between \rx \ and RX\,J0513.9-6951. In addition, a large visual flux
emission from the X-ray heated structures of the binary would explain the absence of
detectable late type features in the optical spectrum. In a $\sim$ 3.8\,d orbit, the 
Roche lobe filling evolved star is expected to have M$_{\rm V}$ $\sim$ 0 (Motch
\cite{m96}).

For stable shell burning, the nuclear luminosity mainly depends on the mass of the
burning envelope and is thus insensitive to short time scale changes in mass
accretion rate (Fujimoto \cite{fujimoto82}). Only for the most massive white
dwarfs which undergo high accretion rates and retain light envelopes can the
nuclear luminosity vary significantly on a time scale of a week.  If the
appearance of the jet is due to a sudden increase of the mass accretion rate onto
the white dwarf, only the much weaker accretion luminosity may vary on short time
scales. Therefore, no large and fast change in bolometric nor optical luminosity is
expected to accompany the jet, in agreement with the V band photometry.

Jet inclinations larger than $\sim$ 60\degree \ seem unlikely as they would imply
outflow velocities in excess of the escape velocity of the most massive white dwarfs
($V_{esc} \ \sim$ 11,000\,km\,s$^{-1}$). On the other hand, if the velocity dispersion
is mainly of geometric origin then the shape of the \Halp \ blue component profile
implies $i$ $\leq$ 29\degree .  Since the jet is likely to be emitted perpendicularly
to the plane of the accretion disc (Becker et al. \cite{becker98}) it is probable that
\rx \ is seen at low inclination angles, consistent with the lack of detected X-ray
eclipses. As for \rxa \ (Becker et al. \cite{becker98}), such low inclinations may be
incompatible with the relatively large amplitude of the photometric light curve.
However, possible jet precession in \rx \ does not allow to draw definite conclusions.

As a whole, the high effective X-ray temperature, small source radius and large jet
velocity hint at a massive white dwarf, which may be close to the Chandrasekhar
limit.

\begin{acknowledgements}
I thank M. Pakull and E. Janot-Pacheco for discussions and comments on an 
early version of this paper and I am grateful to the referee for suggesting 
several valuable improvements.
\end{acknowledgements}

\end{document}